\documentclass[showpacs,prl,superscriptaddress,twocolumn]{revtex4}
\usepackage{amsmath}
\usepackage{amssymb}
\usepackage{amsfonts}
\usepackage{graphicx,bm}
\usepackage{dcolumn}
\usepackage[colorlinks=true]{hyperref}
\usepackage{epsf}
\usepackage{enumerate}
\usepackage{hhline}
\usepackage{array}
\usepackage{tabularx}
\usepackage{subfigure}

\newcommand{\be}{\begin{equation}}
\newcommand{\ee}{\end{equation}}
\newcommand{\bea}{\begin{eqnarray}}
\newcommand{\eea}{\end{eqnarray}}
\newcommand{\beaa}{\begin{eqnarray*}}
\newcommand{\eeaa}{\end{eqnarray*}}

\newcommand{\nn}{\nonumber \\}
\newcommand{\e}{\mathrm{e}}

\def\be{\begin{equation}}
\def\ee{\end{equation}}
\def\bea{\begin{eqnarray}}
\def\eea{\end{eqnarray}}

\begin{document}
\title{Variational principle and boundary terms in gravity {\it \`a la} Palatini}
\author{Diego S\'aez-Chill\'on G\'omez}
\email{diego.saez@uva.es} 
\affiliation{Department of Theoretical Physics, Atomic and Optics, Campus Miguel Delibes, \\ University of Valladolid UVA, Paseo Bel\'en, 7,
47011 - Valladolid, Spain}

\begin{abstract}
A general $f(\mathcal{R})$ gravitational theory is considered within the Palatini formalism. By applying the variational principle and the usual conditions on the boundary, we show explicitly that a surface term remains such that as in their metric-compatible counterparts, an additional surface term has to be added in the gravitational action, which plays a fundamental role when calculating the entropy of the black hole.
\end{abstract}
%
%
\maketitle
%
%
%
\section{Introduction}
Gravitational theories are described by the spacetime metric that measures lengths and covariant derivatives which provide the way vectors are parallely transported. The construction of a gravitational theory departs from an action integrated over the spacetime volume, where the Lagrangian is given by a scalar formed from the Riemann tensor that depends on the connection and its derivatives, and from invariants of the torsion tensor that accounts for the antisymmetric part of the connection \citep{BeltranJimenez:2018vdo}. As far as one assumes a torsion-free metric compatible connection, the so-called Levi-Civita connection arises and one only cares about variations over the spacetime metric, which leads to the gravitational field equations but also to non-null surface contributions. This is the origin of the so-called Gibbons-Hawking-York (GHY) boundary term that compensates the surface term in General Relativity (GR) for spacelike and timelike boundaries \citep{YorkBound,Gibbons:1976ue}. Generalization to null boundaries has been also realized with the appropriate counter-term to be added in the gravitational action \citep{Parattu:2015gga}. Similar boundary terms are found in scalar-tensor theories and metric $f(R)$ gravities \citep{Nojiri:2000kh,Dyer:2008hb,Guarnizo:2010xr}. In the so-called Palatini formalism one considers the spacetime metric and the connection as independent fields in principle, and then by applying the variational principle, the corresponding field equations for the metric and the connection are obtained which for the class of models bases on the Ricci scalar show that actually the connection is metric compatible with a conformal metric to the spacetime one \citep{Olmo:2011uz}. In addition, for such Ricci scalar based theories with matter independent of the connection, torsion can be removed since just appears as a projective mode \cite{Afonso:2017bxr}. This formalism has some advantages in comparison to metric $f(R)$ theories as the equations remain second order and they have been widely studied in the literature inspired by Born-Infeld electromagnetism to construct regular cosmologies and black holes \citep{BeltranJimenez:2017doy}. \\

Nevertheless, in the Palatini formalism, surface terms are removed by considering Dirichlet boundary conditions on the variations of the connection, such that no boundary terms remain. In this letter, we reformulate the variational principle when applied in the Palatini formalism by assuming that the connection is not the fundamental field but the metric compatible tensor is. Some progress in this sense has been done by considering a Levi-Civita connection compatible to a Lorentz metric together with an auxiliary metric \citep{Goenner:2010tr}. Nevertheless, here by assuming Dirichlet boundary conditions on the variations of the compatible metric, the corresponding non-null boundary terms are obtained for the Palatini formalism and we propose a surface counter-term to be added in the gravitational action. Such new term depends conformally on the geometry of the boundary and will play also a fundamental role in several frameworks as the hamiltonian approach of the theory, as the GHY term plays in GR. Here we obtain the entropy of the Schwarzschild black hole in these theories through the Euclidean semiclassical approach and shown that the entropy is induced by the new surface term and the exact expression can not be obtained without this new boundary term. Moreover, the expression for the entropy coincides with the one calculated previously in the Palatini formalism by using Noether charges \citep{Vollick:2007fh}. \\

Hence, the full gravitational action in the Palatini approach is established which opens completely new scenarios when dealing with these theories in multiple frameworks.

\section{Palatini formalism}
\label{background}
The general gravitational action is given by:
\be
S=S_G+S_m=\frac{1}{2\kappa^2}\int d^4x \sqrt{-g} f(\mathcal{R}) +S_m \ ,
\label{fRaction}
\ee
where the metric of the spacetime is given by the symmetric tensor $g_{\mu\nu}$ and we are assuming that the matter action $S_m$ just depends on the metric and the matter fields, preserving the Equivalence Principle, while the constant $\kappa^2=8\pi G$. The Ricci scalar $\mathcal{R}$ is defined as the contraction of the Ricci tensor with the spacetime metric:
\be
\mathcal{R}=g^{\mu\nu}\mathcal{R}_{\mu\nu}(\Gamma)\ .
\label{RicciScalarr}
\ee
In the Palatini formalism the connection is assumed to be in principle an independent field, which might not be metric compatible with $g_{\mu\nu}$. The Ricci tensor is expressed in terms of the independent connection $\Gamma$ as follows:
\be
\mathcal{R}_{\mu\nu}=\partial_{\lambda}\Gamma^{\lambda}_{\mu\nu}-\partial_{\nu}\Gamma^{\lambda}_{\mu\lambda}+\Gamma^{\lambda}_{\sigma\lambda}\Gamma^{\sigma}_{\mu\nu}-\Gamma^{\lambda}_{\sigma\nu}\Gamma^{\sigma}_{\mu\lambda}\ . 
\label{RicciT}
\ee
Note that the connection might not be symmetric in the two lower indices and consequently  the action might contain a non-null torsion component. However, for the class of theories studied here (\ref{fRaction}), the torsion part can be removed. This can be shown by isolating the symmetric part $C^{\lambda}_{\mu\nu}$ from the anti-symmetric part $S^{\lambda}_{\mu\nu}$ of the connection:
\be
\Gamma^{\lambda}_{\mu\nu}=C^{\lambda}_{\mu\nu}+S^{\lambda}_{\mu\nu}\ .
\label{ConnecSymeAnti}
\ee
Then, by applying the following projective transformation, 
\be
\tilde{\Gamma}^{\lambda}_{\mu\nu}=\Gamma^{\lambda}_{\mu\nu}+\frac{2}{3}\delta_{\nu}^{\lambda}A_{\mu}\ ,
\ee
the corresponding components of the connection (\ref{ConnecSymeAnti}) transform as:
\bea
\tilde{C}^{\lambda}_{\mu\nu}&=&C^{\lambda}_{\mu\nu}+\frac{2}{3}\delta^{\lambda}_{(\nu}A_{\mu)}\ , \nn
\tilde{S}^{\lambda}_{\mu\nu}&=&S^{\lambda}_{\mu\nu}+\frac{2}{3}\delta^{\lambda}_{[\nu}A_{\mu]}\ ,
\eea
where $_{[\ ]}$ and $_{(\ )}$ are the commutator and anticommutator on the indices respectively, and the vector $A_{\mu}=S^{\sigma}_{\sigma\mu}$ is chosen such that $\tilde{S}^{\lambda}_{\lambda\nu}=0$. Finally, the connection (\ref{ConnecSymeAnti}) can be expressed as:
\be
\Gamma^{\lambda}_{\mu\nu}=\tilde{C}^{\lambda}_{\mu\nu}-\frac{2}{3}\delta^{\lambda}_{\nu}A_{\mu}\ .
\label{ConnecSymeAnti2}
\ee
And the Ricci tensor is split into its symmetric and antisymmetric parts as follows \cite{Olmo:2011uz}:
\be
\mathcal{R}_{\mu\nu}(\Gamma)=\mathcal{R}_{\mu\nu}(\tilde{C})-\frac{4}{3}\partial_{[\mu}A_{\nu]}\ ,
\label{RicciSymmAnti}
\ee
Then, torsion just enters as a projective mode in the Ricci tensor, such that theories of gravity constructed on the symmetric part of the Ricci scalar and on matter not coupled to the connection, are consequently torsionless  \citep{Afonso:2017bxr}. In addition, theories based on the Ricci scalar (\ref{RicciScalarr}) will remain also torsionless, as the contraction with the symmetric metric leads to:
\be
\mathcal{R}(\Gamma)=g^{\mu\nu}\mathcal{R}_{\mu\nu}(\Gamma)=g^{\mu\nu}\mathcal{R}_{\mu\nu}(\tilde{C})=\mathcal{R}(\tilde{C})\ .
\label{RicciScalarr2}
\ee
This is the so-called projective invariance of the scalar curvature, such that from here on it is justified to remove the torsion part from the connection $\Gamma$ and consider it symmetric in the lower indices.\\

Then, variations of the gravitational action (\ref{fRaction}) consist on variations over the metric and over the connection:
\bea
\delta S_G &=&\frac{1}{2\kappa^2}\int d^4x \sqrt{-g}\left[f_{\mathcal{R}}\mathcal{R}_{\mu\nu}-\frac{1}{2}g_{\mu\nu} f(\mathcal{R})\right]\delta g^{\mu\nu}\nn
&+&\frac{1}{2\kappa^2}\int d^4x \sqrt{-g}f_{\mathcal{R}}g^{\mu\nu}\delta\mathcal{R}_{\mu\nu}\ ,
\label{variations1}
\eea
where $f_{\mathcal{R}}\equiv\frac{df}{d\mathcal{R}}$. Hence, variations with respect to the metric $g_{\mu\nu}$ leads to the following set of equations: 
\bea
  f_{\mathcal{R}}\mathcal{R}_{\mu\nu}-\frac{1}{2}g_{\mu\nu}f=\kappa^2 T_{\mu\nu}\ ,
  \label{Fieldeqs}
  \eea
where $T_{\mu\nu}=-\frac{2}{\sqrt{-g}}\frac{\delta S_m}{\delta g^{\mu\nu}}$ is the energy-momentum tensor. The issue arises when dealing with variations of the gravitational action with respect to the connection. The variation of the Ricci tensor $\delta \mathcal{R}_{\mu\nu}$ is given by:
\be
\delta \mathcal{R}_{\mu\nu}=\nabla_{\sigma}\delta\Gamma^{\sigma}_{\nu\mu}-\nabla_{\nu}\delta\Gamma^{\sigma}_{\sigma\mu}\ ,
\ee
where the covariant derivative $\nabla$ is defined in terms of the independent -symmetric- connection $\Gamma$, and the second term in (\ref{variations1}) yields:
\bea
&&\frac{1}{2\kappa^2}\int d^4x \sqrt{-g}f_{\mathcal{R}}g^{\mu\nu}\delta\mathcal{R}_{\mu\nu}=\nn
&=&\frac{1}{2\kappa^2}\int d^4x \sqrt{-g}f_{\mathcal{R}}g^{\mu\nu}\left[\nabla_{\sigma}\delta\Gamma^{\sigma}_{\nu\mu}-\nabla_{\nu}\delta\Gamma^{\sigma}_{\sigma\mu}\right]\ ,
\label{Variations2}
\eea
which after integrating by parts can be decomposed as follows:
\bea
\int d^4x\ \nabla_{\sigma}\left[\sqrt{-g}f_{\mathcal{R}}\left(g^{\mu\nu}\delta\Gamma^{\sigma}_{\nu\mu}-g^{\mu\sigma}\delta\Gamma^{\lambda}_{\lambda\mu}\right)\right]\nn
- \delta\Gamma^{\sigma}_{\nu\mu}\left[\nabla_{\sigma}\left(\sqrt{-g}g^{\mu\nu}f_{\mathcal{R}}\right)-\delta^{\nu}_{\sigma}\nabla_{\lambda}\left(\sqrt{-g}g^{\mu\lambda}f_{\mathcal{R}}\right)\right]\ .
\label{Variations3}
\eea
The first term in (\ref{Variations3}) is a boundary term and will be analysed later, while the second term after some calculations leads to the well known result:
\be
\nabla_{\lambda}\left(\sqrt{-g}f_{\mathcal{R}}g^{\mu\nu}\right)=0\ ,
\label{compatibilityCond}
\ee
which basically states that the connection is compatible with the conformal metric:
\be
q_{\mu\nu}=\Omega^2 g_{\mu\nu}\ , \quad \Omega^2=f_{\mathcal{R}}\ .
\label{conformalTrans}
\ee
And GR is recovered as far as $f_{\mathcal{R}}=1$. Equation (\ref{compatibilityCond}) becomes:
\be
\nabla_{\lambda}\left(\sqrt{-q}q^{\mu\nu}\right)=0\ .
\label{hmetricity}
\ee
Moreover, by taking the trace of equation (\ref{Fieldeqs}) one obtains:
\be
  f_{\mathcal{R}}\mathcal{R}-2f=\kappa^2 T\ .
   \label{traceRT}
   \ee
This is an algebraic equation for the scalar curvature $\mathcal{R}$, such that can be solved as a function of the trace of the energy-momentum tensor $\mathcal{R}=\mathcal{R}(T)$. From the metric compatibility equation (\ref{hmetricity}), one might consider to apply the conformal transformation (\ref{conformalTrans}) on the Ricci tensor:
\bea
\mathcal{R}_{\mu\nu}(q)=R_{\mu\nu}(g)+\frac{4}{\Omega^2}\nabla^{0}_{\mu}\Omega\nabla^{0}_{\nu}\Omega-\frac{2}{\Omega}\nabla^{0}_{\mu}\nabla^{0}_{\nu}\Omega\nn
-g_{\mu\nu}\frac{g^{\rho\sigma}}{\Omega^2}\nabla^{0}_{\rho}\Omega\nabla^{0}_{\sigma}\Omega-g_{\mu\nu}\frac{\Box^{0}\Omega}{\Omega}\ .
\label{ConformalR}
\eea
Here $\nabla^{0}$ is the covariant derivative defined in terms of the Levi-Civita connection compatible with the spacetime metric $g_{\mu\nu}$. Finally, the field equations (\ref{Fieldeqs}) can be expressed as:
\bea
R_{\mu\nu}(g)&-&\frac{1}{2}g_{\mu\nu}R(g)=\frac{\kappa^2}{f_{\mathcal{R}}}T_{\mu\nu}-g_{\mu\nu}\frac{\mathcal{R}f_{\mathcal{R}}-f}{2f_{\mathcal{R}}}\nn
&-&\frac{3}{2f_{\mathcal{R}}^2}\left[\nabla^{0}_{\mu}f_{\mathcal{R}}\nabla^{0}_{\nu}f_{\mathcal{R}}-\frac{1}{2}g_{\nu\mu}\nabla^{0}_{\lambda}f_{\mathcal{R}}\nabla^{0\ \lambda}f_{\mathcal{R}}\right]\nn
&+&\frac{1}{f_{\mathcal{R}}}\left[\nabla^{0}_{\mu}\nabla^{0}_{\nu}f_{\mathcal{R}}-g_{\mu\nu}\Box^{0} f_{\mathcal{R}}\right]\ .
\label{fieldEq2}
\eea
This is the usual approach when dealing with non linear terms in the action within the Palatini formalism, as the equations are now expressed in terms of the spacetime metric in the left hand side while the right hand side depends solely on the energy-momentum tensor and its trace by the relation (\ref{traceRT}). In general this approach allows to study different spacetimes in a simple way. \\

\section{Boundary terms in the Palatini formalism}
\label{boundary}

Turning back to the surface term in (\ref{Variations3}), this is usually considered to be null in the literature just by applying standard Dirichlet conditions on the boundary, i.e. by assuming that variations of the fields are null on the boundary, in this case variations of the connection $\delta\Gamma$. However, variations of the derivatives of the fields are uncomfortable to be killed, what leads in GR to a non-null surface term that is compensated by the so-called Gibbons-Hawking-York boundary term \cite{YorkBound,Gibbons:1976ue}. The misunderstanding when applying the variational principle to non linear actions in the Palatini formalism has to do with assuming the -in principle- arbitrary connection $\Gamma$ as the fundamental field, while this is not, as shown by the metricity condition (\ref{hmetricity}), but it is the  compatible metric $q_{\mu\nu}$ the fundamental field that plays the game. Hence, by taking this assumption, the boundary term in (\ref{Variations3}) leads to:
\be
\frac{1}{2\kappa^2}\int_{\mathcal{M}} d^4x\ \nabla_{\sigma}V^{\sigma}=\frac{1}{2\kappa^2}\int_{\mathcal{M}} d^4x\ \partial_{\sigma}V^{\sigma}
\label{BoundaryIni}
\ee
where
\be
V^{\sigma}=\sqrt{-g}f_{\mathcal{R}}\left(g^{\mu\nu}\delta\Gamma^{\sigma}_{\nu\mu}-g^{\mu\sigma}\delta\Gamma^{\lambda}_{\lambda\mu}\right)\ .
\ee
Note that for any vector $A^{\sigma}$, the tensor density $\sqrt{-g} A^{\sigma}$ leads to $\nabla_{\sigma}(\sqrt{-g} A^{\sigma})=\partial_{\sigma}(\sqrt{-g} A^{\sigma})$, as $\sqrt{-g}$ is a scalar density. Then, by using the Gauss-Stokes theorem, it yields:
\be
\int_{\mathcal{M}} d^4x\ \partial_{\sigma}V^{\sigma}=\int_{\delta\mathcal{M}} d^3x\sqrt{|\gamma|}\epsilon\ n_{\sigma}V^{\sigma}\ ,
\label{Boundary33}
\ee
where $\gamma$ is the induced metric on the boundary $\delta\mathcal{M}$, $n_{\sigma}$ is the unit normal vector to $\delta\mathcal{M}$ and $\epsilon=\pm 1$ depending whether the boundary is spacelike or timelike. By lowering the index in $V$, this can be expressed in terms of the variations of the metric $q_{\mu\nu}$ as follows:
\bea
V_{\sigma}=g_{\sigma\lambda}V^{\lambda}= \Omega^{-2}q_{\sigma\lambda}V^{\lambda}=\nn
=g^{\alpha\beta}\left[\partial_{\beta}(\delta q_{\sigma\alpha})-\partial_{\sigma}(\delta q_{\alpha\beta})\right]\ .
\eea
And the boundary term (\ref{Boundary33}) yields:
\be
\int_{\delta\mathcal{M}} d^3x\sqrt{|\gamma|}\epsilon\ \ n^{\sigma}g^{\alpha\beta}\left[\partial_{\beta}(\delta q_{\sigma\alpha})-\partial_{\sigma}(\delta q_{\alpha\beta})\right]\ .
\label{Boundary34}
\ee
The spacetime metric can be expressed in terms of the induced metric on the boundary as:
\be
g^{\alpha\beta}=\gamma^{\alpha\beta}+\epsilon n^{\alpha}n^{\beta}\ .
\ee
And (\ref{Boundary33}) leads to:
\bea
\int_{\delta\mathcal{M}} d^3x\sqrt{|\gamma|}\epsilon\ \ n^{\sigma}(\gamma^{\alpha\beta}+\epsilon n^{\alpha}n^{\beta})\left[\partial_{\beta}(\delta q_{\sigma\alpha})-\partial_{\sigma}(\delta q_{\alpha\beta})\right]\ .
\label{Boundary36}
\eea
The projections along the normal direction cancel each other, while the first term is the tangential derivative on the boundary of the variation $\delta q_{\sigma\alpha}$, which becomes null just by imposing standard Dirichlet boundary conditions:
\be
\delta q_{\sigma\alpha}\rvert_{\delta\mathcal{M}}=0\ .
\ee
Finally, the boundary term (\ref{BoundaryIni}) leads to:
\bea
-\frac{1}{2\kappa^2}\int_{\delta\mathcal{M}} d^3x\sqrt{|\gamma|}\epsilon\ \gamma^{\alpha\beta}n^{\sigma}\partial_{\sigma}(\delta q_{\alpha\beta})\ .
\label{Boundary35}
\eea
This is the boundary term that remains in the Palatini approach. It may be expressed in a more convenient way by using the induced conformal metric on the boundary:
\be
\tilde{\gamma}_{\alpha\beta}=q_{\alpha\beta}-\epsilon \tilde{n}_{\alpha}\tilde{n}_{\beta}\ .
\ee
where the normal vector $\tilde{n}_{\mu}$ and $\tilde{\gamma}_{\alpha\beta}$ are related to the induced metric $\gamma$ and the normal vector $n_{\mu}$ by the conformal transformation (\ref{conformalTrans}) as follows:
\bea
\tilde{\gamma}_{\alpha\beta}&=&\Omega^2 \gamma_{\alpha\beta}\ , \nn
\tilde{n}_{\alpha}&=&\Omega n_{\alpha}\ .
\eea
Then, the boundary term (\ref{Boundary35}) yields:
\be
-\frac{1}{2\kappa^2}\int_{\delta\mathcal{M}} d^3x\sqrt{|\tilde{\gamma}|}\epsilon\ \tilde{\gamma}^{\alpha\beta}\tilde{n}^{\sigma}\partial_{\sigma}(\delta q_{\alpha\beta})\ .
\ee
This is nothing but the variation of the Gibbons-Hawking-York boundary term of the conformal metric, such that the appropriate term to be added to the gravitational action is given by:
\be
S_B=\frac{1}{\kappa^2}\int_{\delta\mathcal{M}} d^3x\sqrt{|\tilde{\gamma}|}\epsilon\ \mathcal{K}\ ,
\label{BoundaryPalatini}
\ee
where the conformal extrinsic curvature is given by:
\be
\mathcal{K}=\nabla_{\alpha} \tilde{n}^{\alpha}\ ,
 \ee
being the covariant derivative defined as compatible to the metric $q_{\mu\nu}$. This is related to the extrinsic curvature as defined by the covariant derivative of the normal vector to the boundary $n_{\mu}$ by:
\be
\mathcal{K}=\Omega^{-1}K+3\Omega^{-2}n^{\mu}\partial_{\mu}\Omega,\ \text{with}\ K=\nabla^{0}_{\alpha} n^{\alpha}\ ,
\label{Extrinsic_Conf}
\ee
where recall that $\nabla^{0}_{\alpha}$ is the covariant derivative compatible with the spacetime metric $g_{\mu\nu}$. Hence, the corresponding surface term has been obtained within the Palatini formalism. In the next section, we consider a direct application through the Euclidean semiclassical formalism. 

\section{Euclidean approach and thermodynamics of Schwarzschild Black holes}
\label{Euclidean}

Let us consider the Schwarzschild black hole, which can be described by Schwarzschild coordinates as
\be
ds^2=-\left(1-\frac{2GM}{r}\right)dt^2+\left(1-\frac{2GM}{r}\right)^{-1} dr^2 +r^2d\Omega^2\ ,
\label{Schwa}
\ee
where $d\Omega^2$ is the line element of a unit two sphere. This spacetime metric is a solution of the gravitational field equations (\ref{fieldEq2}) as far as $f(\mathcal{R}=0)=0$ and  the algebraic equation (\ref{traceRT}) has the root $\mathcal{R}=0$. The Euclidean approach consists of approximating the partition function in gravity, which is described by the path integral \citep{Gibbons:1976ue}
\be
Z[\beta]=\int d[g]\ d[\Gamma]\ d[\psi]\ \e^{i S}\ .
\ee
Here $S$ is the gravitational action that includes the corresponding matter fields $\psi$. By applying the saddle point approximation, the main contribution to the path integral is given by the classical action of a Euclidean solution:
\be
Z[\beta]=\e^{\beta F}\approx\e^{-S_E}\ ,
\label{Partition2}
\ee
where $F$ is the free energy and $\beta=T^{-1}$ with $T$ the temperature of the system, while the Euclidean action for the Palatini action (\ref{fRaction}) with the surface term (\ref{BoundaryPalatini}) after applying a Wick rotation $t\rightarrow i\tau$ yields:
\bea
&S_E&=S_{EG}+S_{EB}=\nn
&-&\frac{1}{2\kappa^2}\int_{\mathcal{M}} d^4x \sqrt{-g} f(\mathcal{R}) - \frac{1}{\kappa^2}\int_{\delta\mathcal{M}} d^3x\sqrt{|\tilde{\gamma}|} \mathcal{K}\ .
\label{fRaction23}
\eea
In the case of the Schwarzschild spacetime metric (\ref{Schwa}), the only non zero contribution to the Euclidean integral (\ref{fRaction23}) is the surface term, as $\mathcal{R}=0$ and $f(0)=0$. The hypersurface of integration is given by $r=R$ and in order to make the Euclidean integral convergent at infinity, one subtracts the corresponding contribution of the asymptotic flat spacetime \citep{York:1986it}:
\bea
S_E-S_E^0&=&S_{EB}-S_{EB}^0=\nn
&-&\frac{1}{\kappa^2}\int_{\delta\mathcal{M}} d^3x \left[\sqrt{|\tilde{\gamma}|} \mathcal{K}-\sqrt{|\tilde{\gamma}^0|} \mathcal{K}^0\right]\ .
\label{Euclidean1}
\eea
Here $\tilde{\gamma}$ and $\tilde{\gamma}^0$ are the conformal transformation of the induced metrics on the hypersurface $r=R$ for the Schwarzschild spacetime and the Minkowski spacetime respectively:
\bea 
\gamma_{\mu\nu}dx^{\mu}dx^{\nu}&=&\left(1-\frac{2GM}{R}\right)d\tau^2+R^2d\Omega^2\ , \nn
   \gamma^0_{\mu\nu}dx^{\mu}dx^{\nu}&=&d\tau_0^2+R^2d\Omega^2\ .
\eea
As usual the Euclidean time $\tau$ is integrated over the period $\beta=T^{-1}=2\pi/\kappa=8\pi G M$, with $\kappa$ being the surface gravity and $T$ the temperature. To make the lengths equal for the Euclidean time in the Schwarzschild and Minkowski spacetimes, one imposes $\beta^0=\beta (1-\frac{2GM}{R})^{1/2}$. In addition, the corresponding extrinsic curvatures on the hypersurface $r=R$ are given by:
\bea
K^{0}&=&-\frac{2}{R}\ ,\nn
K&=&-\frac{2}{R}-\frac{MG}{R^2\left(1-\frac{2GM}{R}\right)}\ .
\eea
Hence, by using (\ref{Extrinsic_Conf}) the Euclidean integral (\ref{Euclidean1}) leads to:
\be
S_E-S_E^0=f_{\mathcal{R}}\frac{\beta MG}{2}\left(1-\frac{2GM}{R}\right)^{1/2}\ ,
\ee
which in the limit $R\rightarrow \infty$ yields:
\be
S_E-S_E^0=f_{\mathcal{R}}\frac{\beta MG}{2}=f_{\mathcal{R}}\frac{\beta^2}{16\pi G}\ .
\ee
From the free energy in (\ref{Partition2}), one has $\beta F= S_E-S_{E}^0=f_{\mathcal{R}}\frac{\beta^2}{16\pi G}$ and by using the well known thermodynamical relation for the entropy:
\be
\mathcal{S}=\left(\beta\frac{\partial}{\partial\beta}-1\right)\beta F\ ,
\ee
the entropy $\mathcal{S}$ of the Schwarzschild black hole in Palatini $f(\mathcal{R})$ theories is given by:
\be
\mathcal{S}=f_{\mathcal{R}}\frac{\beta^2}{16\pi G}=\frac{f_{\mathcal{R}}}{4G}A\ .
\ee
Here we have used the area of the horizon $A=4\pi r_s^2$ with $r_s=2GM$. Hence, this is the expression for the entropy in the Palatini formalism which has its origin in the surface term obtained in the previous section. This coincides with the expression in the metric formalism for $f(R)$ gravitational theories \citep{delaCruzDombriz:2009et}. Note also that this result was previously obtained through Noether charges in \citep{Vollick:2007fh} and extended in \citep{Bamba:2010kf}, such that both approaches coincide.

\section{Conclusions}
\label{Conclusions}

In this letter, we have shown that the surface term corresponding to a theory of gravity formulated in the Palatini formalism can not be removed but provides a contribution similar to the well known Gibbons-Hawking-York term of General Relativity and other theories including scalar tensor theories \citep{Dyer:2008hb,Guarnizo:2010xr}. Nevertheless, the surface term here depends upon the conformal geometry of the boundary through the conformal metric $q_{\mu\nu}$ which is sensible to the gravitational action itself through the derivative of the action $f_{\mathcal{R}}$. \\

As in metric compatible theories, the boundary term is necessary for multiple analysis, as the hamiltonian formulation of the theory or the calculation of the black hole entropy in Schwarzschild spacetime through the Euclidean semiclassical approach. The latter has been explicitly shown here, where the expression of the entropy matches the one obtained before in the literature by using Noether charges \citep{Vollick:2007fh}. Moreover, the correct expression is obtained only in the case that the boundary term is included in the gravitational action. Otherwise, the Euclidean approach is not complete when dealing with this type of metrics. Next efforts should lie on the formulation of the hamiltonian approach of the theory and the corresponding Arnowitt-Deser-Misner energy.

\section*{Acknowledgments}
I would like to thank Dr. Diego Rubiera-Garc\'ia for useful comments on the paper. DS-CG is supported by the University of Valladolid (Spain).


\begin{thebibliography}{}

\bibitem{BeltranJimenez:2018vdo}
J.~Beltr\'an Jim\'enez, L.~Heisenberg and T.~S.~Koivisto,
JCAP \textbf{08}, 039 (2018)
doi:10.1088/1475-7516/2018/08/039
[arXiv:1803.10185 [gr-qc]].

\bibitem{YorkBound}
J.~W.~York, Jr.,
Phys. Rev. Lett. \textbf{28} 1082 (1972)
doi:10.1103/PhysRevLett.28.1082

\bibitem{Gibbons:1976ue}
G.~W.~Gibbons and S.~W.~Hawking,
Phys. Rev. D \textbf{15}, 2752-2756 (1977)
doi:10.1103/PhysRevD.15.2752


\bibitem{Parattu:2015gga}
K.~Parattu, S.~Chakraborty, B.~R.~Majhi and T.~Padmanabhan,
Gen. Rel. Grav. \textbf{48}, no.7, 94 (2016)
doi:10.1007/s10714-016-2093-7
[arXiv:1501.01053 [gr-qc]].

\bibitem{Nojiri:2000kh}
S.~Nojiri, S.~D.~Odintsov and S.~Ogushi,
Phys. Rev. D \textbf{62}, 124002 (2000)
doi:10.1103/PhysRevD.62.124002
[arXiv:hep-th/0001122 [hep-th]].

\bibitem{Dyer:2008hb}
E.~Dyer and K.~Hinterbichler,
Phys. Rev. D \textbf{79}, 024028 (2009)
doi:10.1103/PhysRevD.79.024028
[arXiv:0809.4033 [gr-qc]].

\bibitem{Guarnizo:2010xr}
A.~Guarnizo, L.~Castaneda and J.~M.~Tejeiro,
Gen. Rel. Grav. \textbf{42}, 2713-2728 (2010)
doi:10.1007/s10714-010-1012-6
[arXiv:1002.0617 [gr-qc]].

\bibitem{Olmo:2011uz}
G.~J.~Olmo,
Int. J. Mod. Phys. D \textbf{20}, 413-462 (2011)
doi:10.1142/S0218271811018925
[arXiv:1101.3864 [gr-qc]].

\bibitem{Afonso:2017bxr}
V.~I.~Afonso, C.~Bejarano, J.~Beltran Jimenez, G.~J.~Olmo and E.~Orazi,
Class. Quant. Grav. \textbf{34}, no.23, 235003 (2017)
doi:10.1088/1361-6382/aa9151
[arXiv:1705.03806 [gr-qc]].

\bibitem{BeltranJimenez:2017doy}
J.~Beltran Jimenez, L.~Heisenberg, G.~J.~Olmo and D.~Rubiera-Garcia,
Phys. Rept. \textbf{727}, 1-129 (2018)
doi:10.1016/j.physrep.2017.11.001
[arXiv:1704.03351 [gr-qc]].

\bibitem{Goenner:2010tr}
H.~F.~M.~Goenner,
Phys. Rev. D \textbf{81}, 124019 (2010)
doi:10.1103/PhysRevD.81.124019
[arXiv:1003.5532 [gr-qc]].

\bibitem{Vollick:2007fh}
D.~N.~Vollick,
Phys. Rev. D \textbf{76}, 124001 (2007)
doi:10.1103/PhysRevD.76.124001
[arXiv:0710.1859 [gr-qc]].


\bibitem{York:1986it}
J.~W.~York, Jr.,
Phys. Rev. D \textbf{33}, 2092-2099 (1986)
doi:10.1103/PhysRevD.33.2092

\bibitem{delaCruzDombriz:2009et}
A.~de la Cruz-Dombriz, A.~Dobado and A.~L.~Maroto,
Phys. Rev. D \textbf{80}, 124011 (2009)
[erratum: Phys. Rev. D \textbf{83}, 029903 (2011)]
doi:10.1103/PhysRevD.80.124011
[arXiv:0907.3872 [gr-qc]].

\bibitem{Bamba:2010kf}
K.~Bamba and C.~Q.~Geng,
JCAP \textbf{06}, 014 (2010)
doi:10.1088/1475-7516/2010/06/014
[arXiv:1005.5234 [gr-qc]].

\end{thebibliography}
\end{document}